\documentstyle[11pt,newpasp,twoside]{article}
\def\edcomment#1{\iffalse\marginpar{\raggedright\sl#1\/}\else\relax\fi}
\reversemarginpar

\begin{document}
\title{Star Formation and Galaxy Environment}
 \author{H.~K.~C.~Yee}
\affil{Department of Astronomy, University of Toronto,
Toronto, Ontario, M5S 3H8, 
Canada}

\begin{abstract}
The dependence of star formation rate  on galaxian environment
is a key issue in the understanding of galaxy formation and
evolution.
However, the study of this subject is
complex and observationally challenging.
This paper reviews some of the current results, drawing mostly
from recent large redshift surveys such the LCRS, the MORPH collaboration,
and the CNOC1 and CNOC2 redshift surveys.
\end{abstract}

\section{Introduction}

Tremendous strides have been made in the last few years in understanding
the average star formation history of the universe from the present to
as far back as 90\% of the age of the universe (e.g., Steidel et al.~1999,
Sawicki et al.~1997).
However, the overall picture remains crude, and is still subject
to many uncertainties and systematics.
Most of the advances have been based on converting the UV luminosity
density in different epochs into an average star formation rate (SFR).
This requires averaging over volume and extrapolating over luminosity.
The universe contains significant large scale structures even at epochs 
as early as $z\sim3$ (Giavalisco et al.~1998), as 
galaxies formed initially in the most massive dark matter halos.  
Voids, filaments, and rich clusters mark very different environments in
which galaxies reside.
To come to a full physical understanding of galaxy formation and
galaxy evolution,
we need to have a more detailed picture, such
as  the history of star formation in different environments.
This is an observationally challenging task, as currently, we have
barely begun to study these issues in the relatively nearby universe.

There are two major observational questions that we ultimately 
would like to answer.
First, how does the SFR depend on the environment?  Second, how does this
dependence evolve with time?
Currently we have some idea of the answers to the
 first question, and little or no information on the second.
Simple physical arguments provide some expectations on the
influence of environments on SFR; however, these are not always
clear cut.
Processes that lower gas content of a galaxy are expected to decrease
SFR; these include, e.g., 
ram-pressure stripping and evaporation in rich environments;
tidal stripping from close encounters of galaxies; and the decrease
of the accretion rate of new gas into a galaxy in rich environments.
On the other hand, similar processes can also serve to {\it increase}
the SFR: e.g., ram-pressure and tidal shocks, and mergers and 
harassment of galaxies in close encounters.

To obtain definitive and quantitative conclusions, 
well-controlled, large samples (in the thousands) of 
galaxies with redshift, multi-color photometry, 
and spectroscopic information are required.
(Perhaps large, robust photometric redshift samples can provide
some needed advancement also.)
Examples of completed redshift surveys that meet these goals include
the LCRS (e.g., Shectman et al.~1996) for $z\sim0.1$, the CNOC1 cluster
redshift survey (Yee et al.~1996)
with 2600 redshifts in fields of galaxy clusters at $0.17<z<0.55$,
and the CNOC2 field galaxy redshift survey (Yee et al.~2000) with
6000 redshifts at $0.1<z<0.6$.

In this review, which is not intended to be complete,
I will look at three issues.
The first is the star formation--environment dependence in the local
universe.
We will then survey the current knowledge of differential cluster-field
evolution, mostly by looking at galaxy populations in rich galaxy
clusters at moderate redshifts.
Finally some conjectures and remarks are made regarding 
the evolution of SFR and its environmental dependence.

\section{The Local Universe}

In the local universe the SFR-environment relationship, in the sense
that SFR is apparently smaller on average in denser environments,  is
well known, if not well understood.
Indirect evidence supporting this correlation comes from the 
combination of the morphology-density relation of galaxies
(e.g., Dressler 1981) and the correlation between H$\alpha$ or [OII]
equivalent widths and morphological types (e.g., Kennicutt 1983).
However, the question that may be more important to our understanding
of SFR and environment is the more complex issue of
whether similar galaxies in different environments have similar SFRs.
The complexity arises not only from having 
to take galaxy morphology, which is observationally difficult
to quantify, into consideration, but also
that apparent morphological classification is inherently intertwined
with star formation rate.
The latter issue has not been seriously considered in any of
the investigations discussed below.
It is parenthetically noted that the best course for correcting
for morphological distribution will likely require excellent
IR band images where episodic star formation will not skew the
morphological classification as much.

\subsection {Results from the LCRS}
The 26,000 redshift LCRS (Shectman et al.~1996) provides a 
local ($z\sim0.1$) galaxy sample sufficiently large to examine
the SFR in different environments.
However, the lack of color data in the LCRS is a drawback.
Hashimoto et al.~(1998) studied the dependence of SFR on
environment using [OII] strengths in the LCRS.
They quantified the environment using a local galaxy density
parameter and attempted to account for the morphological dependence
by classifying galaxies with a concentration parameter.
They further divided their galaxies into ``cluster'' and ``field''.
They found that in general galaxies considered in ``clusters''
have lower SFR, and that
within the two respective global environments galaxies in
higher density regions have a lower emission line fraction, even
when both results are corrected for the concentration 
parameter distribution.

Allam et al.~(1999) used the same data set but defined
the environment as compact and loose groups, and field.
They found a somewhat inconclusive result that while
the fraction of strong star bursts are higher in the field
by a factor of 2 over that in compact clusters/groups, the
distributions of the equivalent width of [OII] are similar
in all three environments.
This, they concluded, may indicate that 
galaxy morphology accounts for most of the differences in SFR.

\section{Differential Cluster-Field Evolution}

Galaxy clusters offered one of the first direct observations
of galaxy evolution (Butcher \& Oemler 1984), in that more distant
clusters evidently have higher blue galaxy fractions.
Most of what we now know regarding differential evolution of
galaxies in different environments stems from comparing the
field and relatively rich clusters at different redshifts, 
given that measuring galaxy environment
is even more difficult in the more distant universe.
Differential cluster-field evolution is an important piece
in the puzzle of the dependence of SFR on environment.
Such information is not only pertinent in our understanding of
galaxy formation and evolution, but is also vital in cosmology,
e.g., in the determination of $\Omega_m$ using Oort's method
(e.g., Carlberg, Yee, \& Ellingson 1997).

\subsection{The Butcher-Oemler Effect}
The Butcher-Oemler (B-O) effect was verified by the spectroscopic
work of Dressler et al. (1992, and references therein).
%; also Couch \& Sharple 1987). 
They also found a significant percentage of a new spectroscopic
class of galaxies in these intermediate redshift clusters -- 
E+A (or the more recent and appropriate name K+A),
which can be interpreted as post-starburst galaxies.
Zabludoff et al.~(1996) searched for K+A galaxies
in the LCRS sample.
They concluded that at low redshift, unlike the result obtained by 
Dressler et al. for the higher-$z$ B-O clusters, the fractions
of K+A galaxies are similar in the field and clusters and
at the less than 1\% level.

High-resolution imaging from HST of the B-O clusters appears
to show a preponderance of spiral galaxies, leading to 
the conclusion by Dressler et al.~(1997) that the blue galaxies
found in the B-O clusters likely turn into the population of S0
galaxies that dominate the centers of low-$z$ clusters.
However, the process by which this occurs is not certain.

\subsection{Recent Large Spectroscopic Surveys: CNOC1 and MORPH}

The CNOC1  redshift survey (Yee et al.~1996),  a
large spectroscopic survey of EMSS clusters with a
redshift range of 0.17 to 0.55, provides one of the best and largest
data sets to investigate the differential evolution of
field and clusters of galaxies.
The survey contains 2600 redshifts, of which 1300 are cluster
galaxies.
The survey is also unique in that it covers galaxies out to 
cluster-centric radii of well over 2 $h^{-1}$ Mpc.
The field galaxy component has been augmented significantly by
the recent completion of the CNOC2 field galaxy redshift
survey (Yee et al.~2000) which contains about 6000 redshifts.
These two sets of data provide self-consistent comparisons  of the
stellar populations in the two environments.

The CNOC1 data set has been investigated using two complementary
approaches.
One is the classical method of measuring line strengths and
indices.
The results are presented in a series of papers by Abraham et 
al.~(1998), Morris et al.~(1999), and  Balogh et al.~(1999).
The CNOC data allow one to make comparisons of the cluster
and field populations at the {\it same} redshift.
The main conclusion from these investigations is that there is
no significant excess star formation or burst activity over
that of field galaxies at similar redshifts, despite the
presence of the B-O effect.
It is shown that the [OII] equivalent width, $W_0($[OII]$)$, 
which is used as
an indicator of SFR, smoothly increases with
cluster-centric radius to the value found in the field.
There is no excess of either the average $W_0($[OII]$)$, or  the
number of galaxies with 
large $W_0($[OII]$)$, at radii as large as 2.5 $h^{-1}$ Mpc.
Balogh et al.~(1999) also found that the fractions of K+A
galaxies (defined by $W_0 ({\rm H}\delta) > 5$\AA,
and $W_0($[OII]$)<5$\AA) in clusters and field at $z\sim0.3$ are
similar (at about 1 to 2\%), and show only a small, but statistically
insignificant, change from that measured in the local universe
from the LCRS by Zabludoff et al.~(1996).
The CNOC1 results suggest that as field galaxies fall into a rich
cluster, their star formation is truncated in gradual processes, 
and they then evolve passively to become members of the red galaxy population.

The CNOC1 results, however, are apparently discrepant with
those obtained by the MORPH collaboration (Dressler et al.~1999;
Poggianti et al.~1999).
The MORPH collaboration found  a much higher fraction of K+A and
starburst (A+em) galaxies in clusters compared to the field.
For example, with a definition of $W_0 ({\rm H}\delta) > 3$\AA~for
K+A galaxies, they found fractions of 21\% and 6\% for clusters and
field, respectively.
The discrepancy with CNOC1 results remains very 
large after adjusting for the 
different definitions.
Balogh et al.~(1999) provide an extensive discussion for possible
explanations for the differences, which include galaxy and cluster
sample selections and measurement methods.

\subsection{Principle Component Analysis of the CNOC1 Sample}

A new approach in studying the populations and star formation
in galaxies is using the technique of Principle Component Analysis 
(PCA, e.g., Connolly et al.~1995), which allows one to derive the relative 
fraction of the stellar populations in galaxies without making discrete
individual measurements such as line indices.
Preliminary results from the CNOC1 survey are reported
in Ellingson et al.~(1999).
Using the PCA method, we can deproject the stellar populations
of  cluster galaxies
into three components: ``cluster-like'', ``field-like'', and
``post-star-formation'' (psf).
From the combined data of 15 clusters, Ellingson et al.~found that
the cluster-like component shows a strong positive gradient towards
the center, while the field-like component shows a 
strong negative gradient.
Furthermore, the components 
match onto the field values at about 2.5$r_{200}$
(where $r_{200}$ is the radius at which the interior average mass density
of the cluster is 200$\rho_c$), or about 3 $h^{-1}$ Mpc, 
 
The PCA decomposition also shows that the field-like and psf components
have a larger spatial extent than that of the cluster-like component.
Separating the CNOC1 sample into a $z>0.3$ (median $z$ of 0.42)
and $z<0.3$ (median $z$ of 0.23) sample, it is found that
the higher $z$ sample has a larger field-like fraction outside of the
central 0.25 $r_{200}$.
This is simply a restatement of the B-O effect, but now using
the field galaxies at the redshifts of the clusters as fiducials.
%Figure 1, taken from Ellingson et al., illustrates the relative 
%luminosity density profiles of the cluster-like
%and field-like components of the CNOC1 clusters in the two redshift
%bins.
Ellingson et al.~also computed the average luminosity profiles of
the cluster-like and field-like components for clusters in the two
redshift bins.
The interesting result is that the red (i.e., cluster-like)
component essentially has an identical profile in the two
epochs, while the blue (i.e., field-like) component for the
higher-$z$ subsample shows
a drop relative to the lower-$z$ clusters while retaining a 
similar profile shape.
This suggests that the B-O effect can be explained most simply
by a change in the field galaxy infall rate over the redshift range,
rather than a change in the physical processes of converting blue
field galaxies into red galaxies over these time scales (although
such explanations are not ruled out).

\section{Groups, Pairs, and Poor Environments}

The dependence of galaxy evolution as a function of environment
other than in rich cluster is an observationally challenging 
subject of study, and
currently we basically know very little in this area.
The primary obstacle is the difficulty in defining the environment
of a galaxy even at moderate redshifts; e.g.,  producing a well-understood
sample of galaxy groups of different richness.

Allington-Smith et al.~(1993) used radio galaxies as markers 
for galaxy groups and clusters of varying richness.
They compared the blue-galaxy fraction ($f_B$) of 14 groups 
of varying richness at $z\sim0.3$
with a larger sample of groups at $z\sim0.05$ from the CfA
redshift survey.
For the CfA groups, it was found that $f_B$ increases from less
than $\sim0.05$ for the rich groups to $\sim 0.3$ over 
a richness range of about 20.
For the higher $z$ groups, they concluded that, although $f_B$ 
increases in general, the color-environment relationship seen
in the CfA sample no longer exists.
However, it appears that their conclusion was dependent on 
only two poor, but apparently red, groups in their high-$z$ sample.
One can just as easily conclude that, based on their Figure 19, the
$f_B$-richness relation for their high-$z$ sample is entirely
consistent with that of their low-$z$ sample (their Figure 16)
shifted blueward by the same amount for all environments. 

The CNOC2 Field Redshift Survey (Yee et al.~2000) is currently
the largest redshift sample at the intermediate 
redshift range of $\sim0.35$.
An  initial report on the evolution of the LF of the galaxy sample
(as a whole) has been given in Lin et al.~(1999).
More detailed analyses of the evolution of galaxies are
underway and it is expected to offer some rudimentary
clues on the evolution of galaxies in different environments.
Very preliminary indications are that the relative evolution
of galaxies in different environments may not be drastically
different, at least for the more luminous galaxies.
This is supported by an analysis of close spectroscopic pairs
(Patton et al.~2000, in preparation; Carlberg et al.~2000a), 
in which the derived 
merger rate show very little evolution up to redshift of 1
(as opposed to a number of earlier studies).
Furthermore, there appears to be little or no evolution in
the clustering length of luminous galaxies over this redshift range, 
leading Carlberg et al.~(2000b) to conclude that the evolution
of $M^*$ galaxies seen over the intermediate redshift range is
not driven by a change in environment.
Also, a very preliminary analysis of the SFR 
(as indicated by the spectral energy distribution) 
of the CNOC2 sample as a function
of local galaxy space density (as measured by the richness parameter
 $B_{gc}$ [see Yee \& L\'opez-Cruz 1999])  shows that the 
changes over redshift are similar over a span of a factor of 10 in
richness.

\section{Summary Remarks}

The interplay between galaxian environment and star formation is clearly
a key issue in the understanding of galaxy formation and evolution.
It is also a difficult and complex problem and requires extensive
observational resources to make definitive progress.
There are clear indications of the effect of the environment when we
compare extreme cases of the cores of rich clusters and the field.
However, the environmental effect on the rate of galaxy evolution
may be subtle, in that over a less extreme range definitive
differences in relative evolution have not been observed conclusively.
Some of the most intriguing and useful work in this area may come from
the direct observations of the effects of a changing environment on star
formation and galaxy evolution as found
in the infalling galaxies at the outskirts of rich clusters.
A thorough investigation of the phenomenon covering  a large span of
cluster-centric radius and over a range of redshift may provide
a crucial piece of the puzzle.
More detailed investigations into the star formation properties
of galaxies in clusters, such as those  
carried out by Moss \& Whittle
(2000, in these proceedings) using H$\alpha$ narrow-band imaging,
has the possibility of providing physical details
on how the star formation process is affected when a galaxy falls
into a cluster.

\end{document}